\documentclass[logo,url,11pt,a4paper]{ETHpaper}

\usepackage{graphicx}        % Figures
\usepackage[sort&compress]{natbib}     % Bibliography
\usepackage{amsfonts}
\usepackage{array}
\usepackage{amsthm}
\usepackage{amsmath}
\usepackage{appendix}    
\usepackage{url}
\usepackage{savesym}
\usepackage{amsmath}
\usepackage{amsthm}
\savesymbol{iint}
\usepackage{txfonts}
\usepackage{overpic}
\usepackage{mathtools}
\begin{document}

\newcommand{\mean}[1]{\left\langle #1 \right\rangle}
\newcommand{\abs}[1]{\left| #1 \right|}
\newcommand{\comment}[1]{\vspace{0.5cm}\textit{#1}\vspace{0.5cm}}

\title{Credit Default Swaps Drawup Networks: Too Tied To Be Stable?}

 \author{Rahul Kaushik and Stefano Battiston}
 \address{Chair of Systems Design, ETH Zurich, Switzerland\\
  \texttt{rkaushik@ethz.ch, sbattiston@ethz.ch}}

   %\url{sbattiston@ethz.ch}

%\date{August 29th, 2011}

%\authoralternative{Author names omitted for double-blind review}

%\reference{Revised version (05 May 2010) for re-submission to \\[-0.5ex] \textit{Information Systems and e-Business Management}.}

\date{\today}

% If you would like to have alternative an title or author in the framing
% (header and footer) of the document, specify them with
\titlealternative{Kaushik and Battiston, CDS Networks}
% and
% \authoralternative{}

% Reference to the article:

% Conventions: if quotable --> \textsl{Name of Journal} \textbf{Volume} (2006) 999-999
%              else        --> (submitted)

%\reference{\emph{In}: A. Editor1, B. Editor2, C. Editor3 (Eds.):
%  \textsl{Booktitle}, Address: Publisher, 2006, pp. 999-999}

% URL displayed below the logo:

\www{\url{http://www.sg.ethz.ch}}

% If not using the logo, append
%
% See \url{http://www.sg.ethz.ch} for more information.
%
% to the reference in a separate line.

\makeframing
\maketitle

\begin{abstract}
  We analyse time series of CDS spreads for a set of major US and European
  institutions on a period overlapping the recent financial crisis. We
  extend the existing methodology of \emph{$\varepsilon$-drawdowns} to
  the one of \emph{joint $\varepsilon$-drawups}, in order to estimate
  the conditional probabilities of abrupt co-movements among
  spreads. We correct for randomness and for finite size effects and
  we find significant probability of joint drawups for certain pairs
  of CDS.  We also find significant probability of trend
  reinforcement, i.e. drawups in a given CDS followed by drawups in
  the same CDS. Finally, we take the matrix of probability of joint
  drawups as an estimate of the network of financial dependencies
  among institutions. We then carry out a network analysis that
  provides insights into the role of systemically important financial
  institutions.
\end{abstract}

\section*{Introduction}
\label{Introduction}

%TO BE INCORPORATED FROM THE CDSPaper_test.tex- DONE (with corrections, I like it the way it is now)
Within the field of complex networks \citep{Caldarelli2007Scale-FreeNetworksComplex}, the investigation of financial networks is currently one of the emerging avenues \citep{Schweitzer2009economic_networks}, also in view of the on-going global financial crisis. Financial contagion on networks \citep{battiston2012cascades} differs in some important respects from the well-known processes of epidemics spreading \citep{barrat2008dpc,Pastor-Satorras.Vespignani2001EpidemicSpreadingin}. It also bears similarities to epidemics spreading, among which, the fact that the topological structure of the network plays a crucial role in the collective dynamics and therefore in emergence of systemic risk.
% One of the major challenges in the investigation of financial networks and systemic risk is that the detailed information on the exposures among the main actors are not publicly disclosed, due to confidentiality and strategic issues. As a result, although various regulatory bodies are pushing in this direction, no global repository for financial network data exists at the moment. Empirical works have covered only few national networks which only in some cases reflect counterparty risk \citep{cajueiro2008brazilian_interbank, Cont2010network_structure,Upper.Worms2004EstimatingBilateralExposures,Soramaki2007topology,iori2008overnight,Mistrulli2007assessing}.
% Our paper takes on this challenge by aiming to provide some insights into the structure of dependencies among financial institutions participating to the Credit Default Swaps (CDS) market, based on the analysis of $\varepsilon$-drawups, \citep{johansen2003characterization,sornette2006predictability} in the respective CDS's spread time series. This is the first attempt to reconstruct the network of CDS contracts, despite the growing volume in size and systemic relevance of this market in the recent years ( see section 2, SI).
% %\ref{sec:backgr-inform-cds}).
There is a body of work on networks reconstructed from correlations among equity prices or return time series \citep{Bonanno2003correlation,kullmann2002time,garas2008structural}. The analysis of the minimum spanning tree provides insights into the classification of stocks and the level of correlation depending on the market phase. Correlation analysis suffers, however, from some important limitations, the main one being that zero correlation between two series does not imply that they are independent (only the inverse is true, see more details in section 4 in the SI).  To overcome these limitations, here we utilise a method based on the detection of \textit{joint} $\varepsilon$-drawup, which allows us to estimate the probability that two series exhibit a co-movement. Moreover, in contrast to equity, CDS prices reflect %directly
the default probability of the reference entity and thus the network constructed from CDS prices are more relevant in studying the propagation of default risk.

Our approach can be applied to construct networks of dependencies in other financial markets. In general, it applies to all domains of networks in which links are, for any reason, unobservable but the dynamics of the nodes reflect the dependency structure. To summarise, the contributions of the paper are the following. First, we build on the $\varepsilon$-drawup method \citep{sornette2006predictability} to estimate the probability of \emph{joint $\varepsilon$-drawups}, which are essentially a particular type of co-movements across time series. Based on this, we estimate the level of the so-called interdependence and trend reinforcement in the system across different phases of the market.  In addition, we find them to be relevant indicators for the emergence of systemic risk, as was also predicted by previous theoretical work. Second, we construct a network of interdependencies among institutions and we introduce two novel centrality measures that allow the identification of systemically important nodes in the network. Our approach enables the disentanglement of a structure that is, only apparently, very homogenous. It also allows us to track the role of nodes evolving in time.

\section{Results}
\label{sec:results}
\begin{figure}[htb!] 
\includegraphics[width=0.97\textwidth]{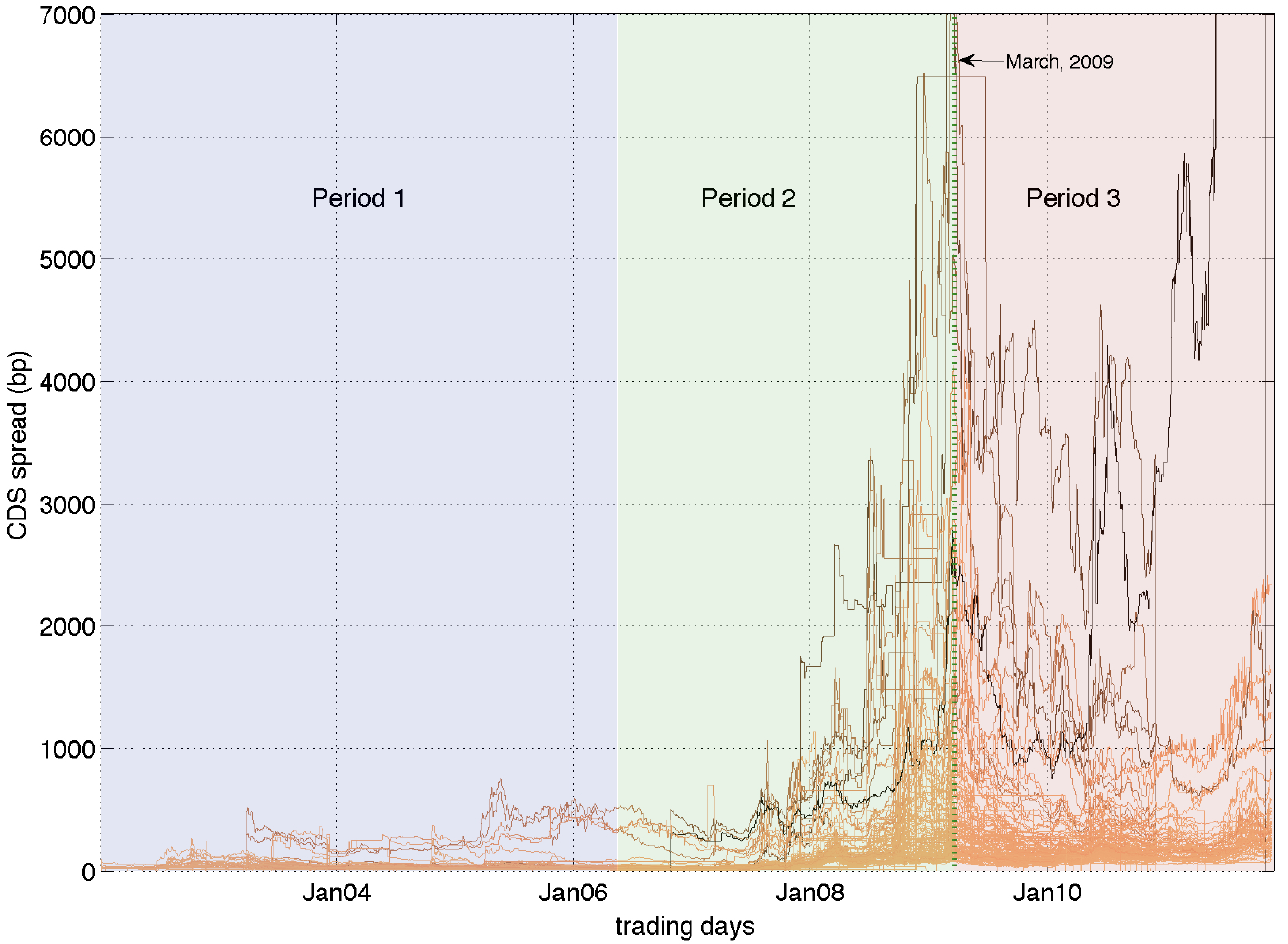} 
  \caption{
\textbf{Time series of credit default swaps throughout the credit crisis}. A plot of the CDS spread time series covering the financial crisis of 2008.   The data ranges from January 2002 to December 2011. We can observe three market phases. Most CDS spreads peak around March 2009.  The CDS prices are quoted in basis points (bp).
\label{fig:traces}
} 
\end{figure}

\textit{Credit Default Swaps} (CDS's) are financial derivatives instruments in which the seller provides the buyer protection against a credit event of a reference entity (see section 2 of the SI).  We have analysed the time series data of  CDS prices, or spreads,  of top US and European financial institutions from January 2002 `till December 2011. We observe three different phases in the CDS prices, corresponding to the periods before, during, and after the credit crisis of 2008: a normal phase (period 1); a volatile and upwards trending phase (period 2), and a volatile and downwards trending phase (period 3) (see figure \ref{fig:traces}).  Accordingly, we divide the data into three parts.  For each institution's time series we detect the so called $\varepsilon$-drawup's. An $\varepsilon$-drawup is an extension of the notion of an $\varepsilon$-drawdown  (\citet{johansen2003characterization}). It refers to a persistent upward movement in a time series until a peak has been reached, after which the time series declines (or, has a ``correction'') by more than an amplitude $\varepsilon$ (see figure \ref{fig:cds_traces_drawup_det}(b)).  Since the CDS spread represents the cost of insurance, an $\varepsilon$-drawup signifies an increase in the default probability of that institution, as perceived by the market.  When market participants buy and sell insurance on each other, their financial performances can become interdependent (see section 3 of the SI).   

Whilst interdependence can be seen as a form of risk diversification which decreases individual risk, previous work \citep{battiston2009liaisons} has demonstrated that high interdependence leads, instead, to higher systemic risk when coupled to a so-called trend reinforcement.

\textbf{Interdependence and trend reinforcement}.

\begin{figure}[htb!] 
\begin{centering}
   \includegraphics[width=0.97\textwidth]{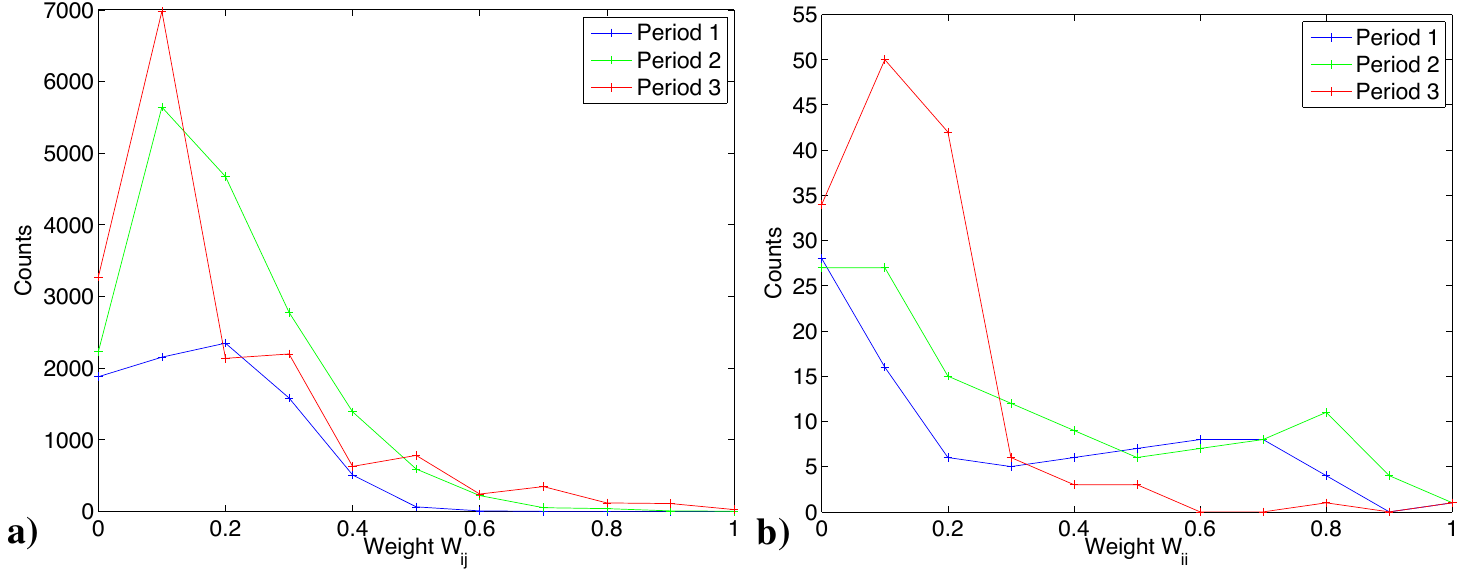} 
    \caption{
  \textbf{a)}The distribution of non-zero values of interdependence $W_{ij}$ across the three periods. The counts in periods 2 and 3 are higher than in period 1.  In addition, during periods 2 and 3 the distributions of $W_{ij}$ have longer tails compared to period 1.  
  \textbf{b)} The distribution of non-zero values of trend reinforcement $W_{ii}$ across the three periods.  } 
 \end{centering} 
 \label{fig:wt_dist_impact_cent}
\end{figure}

In our context, trend reinforcement refers to the tendency of an $\varepsilon$-drawup to be followed by another $\varepsilon$-drawup in the same time series. Interdependence refers, in contrast, to the existence of co-movements between two different time series (i.e. an $\varepsilon$-drawup followed by another one in a different time series). Here, we take the frequency of $\varepsilon$-drawup's in $i$ as an estimate of the probability $P_i$ that security $i$ has a $\varepsilon$-drawup.  Similarly, for the frequency of \textit{joint} $\varepsilon$-drawup's we estimate $P_{ij}$, i.e. probability that $j$ experiences a $\varepsilon$-drawup given that $i$ experiences a $\varepsilon$-drawup. We also account for a time lag $\tau= 0,1,2,3$ days between the drawup's. The expected probability of joint drawup's in the case of two statistically independent time series is $P_{ij}=P_iP_j$. Therefore, we take as an estimate of interdependence between two financial institutions, the deviation from such a case, i.e. $W_{ij}=P_{ij}-P_iP_j$.  Finally, in order to account for finite size effects, we consider only the values of $W_{ij}$ that cannot be rejected based on a permutation test and we reset to zero all the other values (see Methods).  We show the distribution of $W_{ij}$'s and $W_{ii}$'s (Fig. \ref{fig:wt_dist_impact_cent}a), b).
Notice that, because of the time lag, $W_{ij}$ is not a symmetric matrix. A positive value of $W_{ij}$ does not prove any causality between the movements of $i$ and $j$, but measures the dependence of $j$ due from $i$, or equivalently, the impact of $i$ on $j$.  Accordingly, the histograms (Fig. \ref{fig:wt_dist_impact_cent}a), b) only show positive values of $W_{ij}$'s and $W_{ii}$'s.  

Analogously, we take as an estimate of the trend reinforcement for institution $i$ the deviation: $W_{ii}=P_{ii}-P_iP_i$ and we treat it as above.

We find a significant level of trend reinforcement (i.e. which cannot be accounted by chance or finite size) in about 50\%, 72\% and 80\% of the nodes (respectively in period 1, 2 and 3). This is reflected in the histograms of $W_{ii}$ (Fig. \ref{fig:wt_dist_impact_cent}b). As we can see, the curve for period 2 and 3 is mostly above the one for period 1. This means that the number of nodes with a significant level of trend reinforcement increases when the market moves from the first phase to more volatile phases. We also find a significant level of interdependence in 54\%, 78\% and 77\% of pairs of nodes in period 1, 2, 3, respectively. The histograms of $W_{ij}$ (see Fig. \ref{fig:wt_dist_impact_cent}b) show that periods 2 and 3 are characterised by higher frequencies. In fact, 20\% of pairs in period 2 and 19\% pairs of nodes in period 3 exhibit values of $W_{ij}$ greater than the mean plus one standard deviation of period 1. Moreover, while in period 1, nearly all values of $W_{ij}$ are smaller than 0.5, in period 2 and 3 there is a tail extending up to 1. 

These findings show that interdependence and trend reinforcement are indeed present in an important market such as the one for CDS's. Moreover, trend reinforcement increases from period 1 to period 2, and even more so does interdependence. Before drawing the implications of these findings in terms of systemic risk, we proceed to a network analysis of the structure of interdependencies among institutions.

\textbf{Network analysis and systemic risk}.
The CDS market can be naturally mapped into a directed and weighted network in which nodes represent institutions and edges represent interdependencies among institutions. More precisely, whenever $W_{ij}>0$, we assign a weighted edge with value $W_{ij}$ from institution $i$ to $j$. Since $W_{ij}$ is the probability that conditional to $i$ experiencing a draw-up, $j$ also experiences a draw-up with a time lag $\tau$, it follows that the stronger the edge, the stronger the impact that $i$ has on $j$.

When looking at the properties of connectedness of the network,  we find a significant number of disconnected nodes in all three periods (81, 39, 39, respectively). Remarkably, the rest of the nodes form only one strongly-connected component (LSCC, see Methods) encompassing, respectively, 95, 137, 137 nodes in period 1, 2 and 3. The density of links (i.e. the number of links over the number of possible links) in the LSCC is high in all the three periods: 0.98,0.97, 0.97.
% TODO: add/fix numbers above
% DONE
This is reflected also in the average out degree in the LSCC's across the three periods, which is $90 \pm 8$, $129 \pm 11$, and $123 \pm 19$. Finally, the average path length within the LSCC's is 1.04, 1.05 and 1.2, meaning that almost all the nodes in the LSCC are first neighbours to each other. In such a structure, each node has a direct impact on all the other nodes, and each of these has a further impact on all the others. Intuitively, this finding suggests that the financial distress at one node in the SCC can quickly propagate to all the other nodes in the LSCC and keeps reverberating through the many connections.  

Indeed, recent works on systemic risk in financial networks have shown that the number of links play an ambiguous role. Few links are functional to diversify the individual risk. However, too many links generate systemic risk. This holds in presence of mechanisms that either amplify the distress (such as in the case of contagion), or simply increase the persistency of the distress in time, such as trend reinforcement \citep{battiston2009liaisons}. As we have seen, the CDS market exhibits a core of more than 100 nodes, that is almost a fully connected graph (i.e. with maximal degree) and where in many cases links represent strong interdependencies. According to the theoretical results mentioned earlier, in such a situation, even small levels of amplification can make the whole system very unstable. Notice that moving from period 1 to period 2, the values of most CDS's raised dramatically, in many cases by one order of magnitude (see Fig. \ref{fig:traces}).  Looking at individual institutions it is clear that their risk of default had become very high.

However, previous findings suggest that, in such a situation, the default of a few institutions would have triggered a systemic default. Indeed, it is generally thought that without the massive intervention of the \emph{Federal Reserve} (FED) through various emergency programs that lasted from the fall of 2008 until the summer of 2009, \citep{BloombergDataDic2011}

Therefore, monitoring trend reinforcement and interdependence together with the individual riskiness of the participants seems to provide a valuable assessment of the level of systemic risk in a market.

\textbf{Centrality}. In order to gain insights into the systemic importance of specific nodes in the network, we proceed to investigate their centrality. We obviously focus only on the strongly connected component since the other nodes are isolated.

The out- and in-degree of a node are the simplest measures of centrality that hold a valuable interpretation here: A high out-degree represents the ability of a node to affect many neighbours when it experiences a draw-up; a high in-degree corresponds to a node being affected by many nodes. Since the network is almost a complete graph, based on the out-degree, all nodes are equally systemically important and equally affected by the others. Interestingly, even the out- and in-degree strength of a node (i.e. the sum of the weights in the outgoing and incoming links), yield similar conclusions.

As an alternative approach, based on the notion of feedback centrality,  for each node $i$ we introduce a novel measure of systemic importance, called \emph{impacting centrality} and denoted as $b_{i}$. The measure takes into account, in a recursive way, the fact that a node is more systemically important if it impacts many systemically important nodes (see Methods). Symmetrically, we also introduce a measure of systemic vulnerability, called \textit{impacted centrality} of a node $i$ and denoted as $c_{i}$.  The latter measure captures the idea that a node is more vulnerable if it has strong dependencies from many nodes which are in turn heavily impacted by the whole network -- Recall that we are only considering upward movements in CDS spreads, i.e. an increase in the fragility of the underlying reference firm. 

\begin{figure}[htb!]
\begin{centering}
\includegraphics[width=0.97\textwidth]{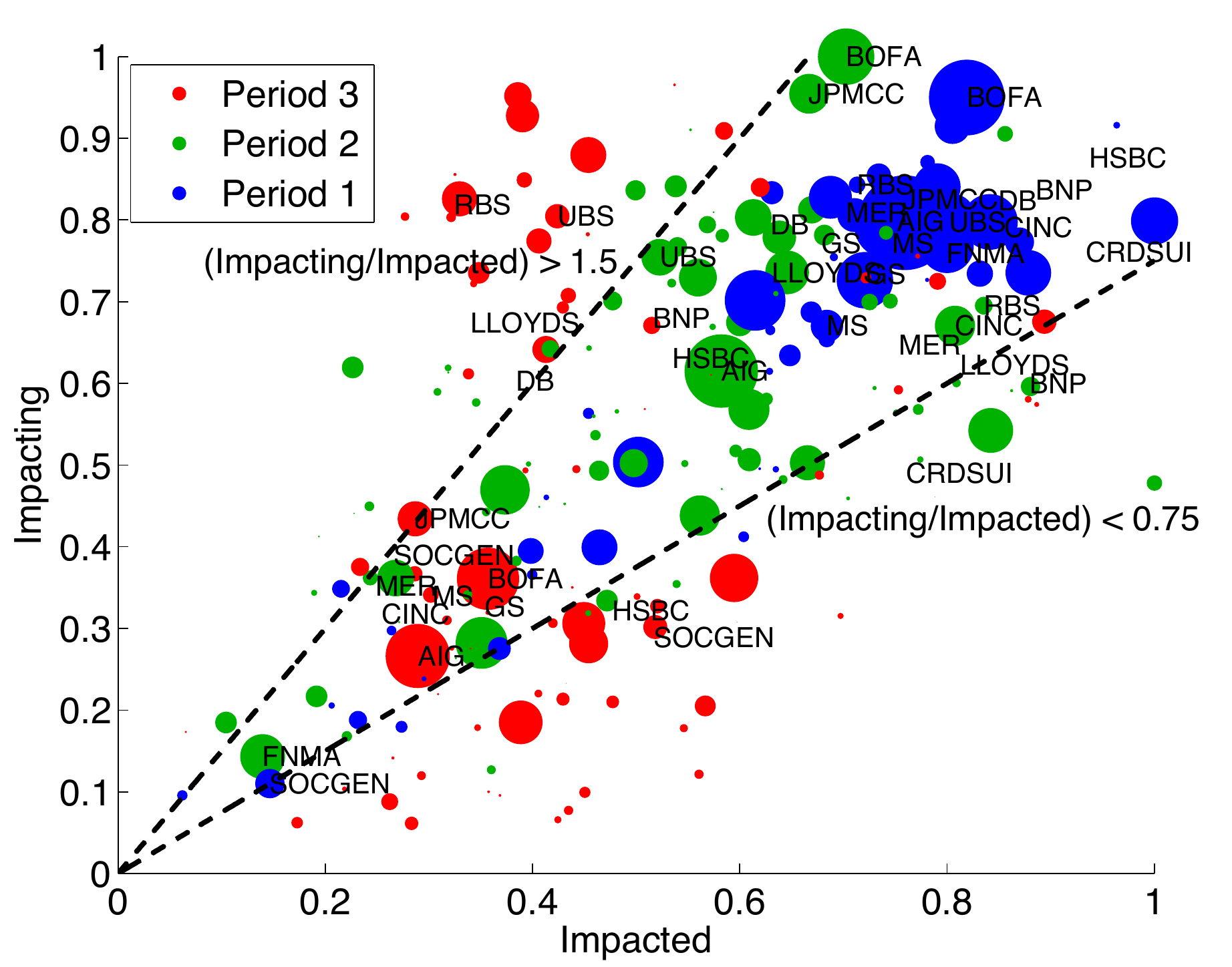} 
\caption{ \textbf{Scatter plot of impacting versus impacted centrality}. Each institution in the CDS market is represented by three dots depending on the period (blue, green, red refers to period 1, 2, 3, respectively). The size of each node is determined by the leverage of a financial institution.  It can be seen that, while in period 1 most institutions are located between the two dotted lines, in period 2 and 3 many of them move to the top and bottom region. This means that ratio between the two centrality measures varies with the market phase. Few institutions of interest are labelled. For example, Bank of America (BOFA) remains in the same region across the three periods. With reference to the subsequent bow-tie construction used in Fig. \ref{fig:bow-tie}: The scatter plot is divided into 3 regions. Nodes in the region above the line $r_{i}>3/2$ correspond to the IN.  Nodes in the region $2/3 < r_{i}<3/2$  correspond to the SCC.  Nodes in the region $r_{i}<2/3$ correspond to the OUT. } 
\end{centering} 
 \label{fig:scatter_b_versus_c}
 
\end{figure}

In both cases, the values are normalised between 0 and 1. Remarkably, in stark contrast to in- and out-degree, the values of these two centrality measures are broadly spread across the range $[0, \, 1]$ (see Fig. \ref{fig:scatter_b_versus_c}, also SI, Fig. 6a \& 7a).

If we focus on the ratio $r_{i}=\frac{b_{i}}{c_{i}}$, in the scatter plot of Fig. \ref{fig:scatter_b_versus_c}, it is possible to identify three regions (above, between and below the dotted lines), corresponding to three different roles. Nodes are located in the top region if they have a value $r_{i}>3/2$, meaning that they impact the network 1.5 times more than they are impacted by it. Symmetrically, nodes are in the bottom region if $r_{i}<2/3$. Finally, nodes that appear in the middle region are those that impact and are impacted by the network in a comparable manner.

A scatter plot of impacting vs impacted centrality in each period is shown in Fig. \ref{fig:scatter_b_versus_c}, where the size of each node reflects the average (relative) debt held by a financial institution during that period.  
According to the above classification, in period 1 most institutions are located in the middle region, implying that the most vulnerable nodes are also the most systemically important.  Moreover, the nodes with the largest debt also happen to be the most systemically important and vulnerable, as they are located in the top right corner.

In period 2 and 3 most institutions progressively move to the top and the bottom regions.  For instance, in period 3 (red bubbles) we can observe one cluster nodes with large debt in the top centre corner of the plot.  This reflects that these nodes have a high impacting centrality, but only an intermediate level of impacted centrality.  A possible explanation is that these institutions managed to reduce their exposure to their counterparties from period 2 to period 3.  There is also a cluster of nodes with large debts located in the bottom centre of the plot.  These nodes possibly correspond to firms that not only managed to reduce their exposure, but also became less systemically important.

We observe that there are many nodes in the network that not only have a high impacting centrality, but also a high impacted centrality.  
In addition, by adjusting the size of each node based on the average debt held by a financial institution a clearer picture emerges.  

Unlike period 2, financial institutions with the largest debt's were in the top right of  Fig. \ref{fig:scatter_b_versus_c}.  The top right corner of Fig. \ref{fig:scatter_b_versus_c} represents firms that are just as vulnerable as they impact the others in the network.  Since, the size of the debt of such institutions is the largest, a small percentage change in their debt would cause a large change in the distress of the debt issued by other firms in the network. We find that in contrast to the evolution of CDS spreads in period 1, the impacting and impacted centralities along-with average debt of the institutions implied ripe conditions for greater levels of systemic risk.  We find that firms with high average debt that are mostly being impacted are important from a systemic risk perspective.  However from a financial regulator's perspective it is also important to track firms, based on their size, impacting, and impacted centrality, that have the ability to cause distress to a large subset of the market.

From a systemic risk perspective it is essential to study nodes that are prone to distress; however, from a policymakers perspective it is also vital to monitor nodes that are not only prone to distress, but that also have a high impacting centrality as distress in such nodes could lead to a systemic collapse.  
  
\textbf{Bow-tie extraction}. 
In order to emphasise the 3 different roles suggested by this finding, we carry out the following filtering of the links. In each period, for nodes located in the top region, we remove all their incoming links. Symmetrically, for those in the bottom region, we remove all the outgoing links. Since the initial network is strongly connected, in this way.  We, by construction, obtain a bow-tie structure (see Methods). The position of a node in the bow-tie is related to its systemic importance. Indeed, the IN, SCC and OUT component of the bow-tie correspond to the top, middle and bottom regions of Fig. \ref{fig:scatter_b_versus_c}, respectively (e.g. the nodes in the IN are those that impact the network more than they are impacted).  

Note that the \emph{bow-tie} structure is constructed based on the choice of impacting-impacted centrality, i.e. nodes with $r_{i}>3/2$ are in the IN, nodes with $2/3<r_{i}<3/2$ are in the SCC, and nodes with $r_{i}>2/3$ are in the OUT.  In fact, for any $\delta >0$, where $\delta \in (0, 1)$.  The lines $1-\delta$ and $1+\delta$ would separate the nodes into three regions.  Thus, the choice of $\delta$ is based on the level of impacting-impacted centrality that is of interest, see SI for more visualisations.  In addition, if $W$ is a directed SCC, and one truncates all incoming links of nodes with $r_{i}<1-\delta$, and all outgoing links for nodes with $r_{i}>1+\delta$.  Then, it is not always the case that the filtered $W$ has a non-trivial SCC, see SI for more details.

\begin{figure}[htb!]
\begin{centering}
\includegraphics[width=0.7\textwidth]{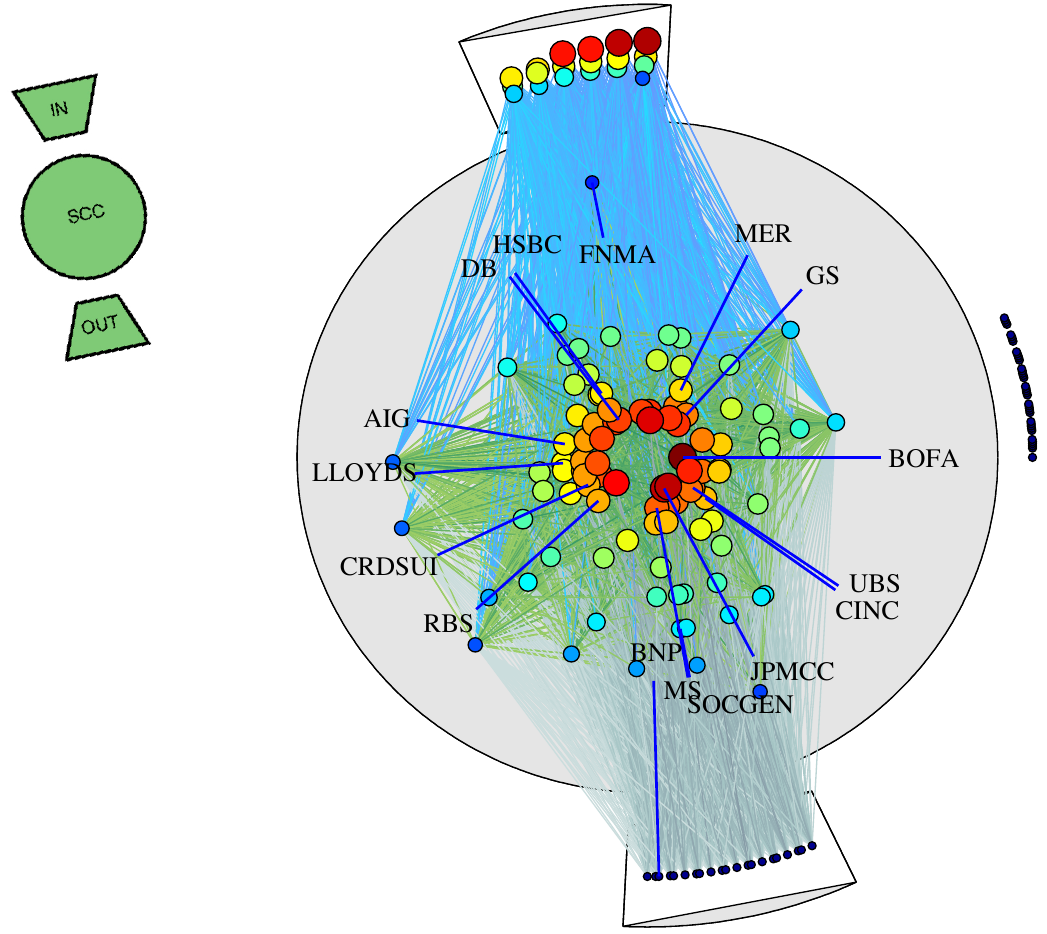} 
\caption{\textbf{The network of the CDS reference entities from period 2}. Each of the nodes represents a financial institution. Outgoing links from nodes that are in the top, or the IN of the bow-tie structure represent the estimated potential impact of a financial institution to its neighbours (see Methods). The nodes in the SCC are placed within a circle of radius one and centred at the origin. The distance of each node from the centre is $1-$Impacting centrality.  The angle increases linearly from  $0$ to $2 \pi$. Thus, the closer a node is to the centre the higher is impacted-impacting centrality.  Similarly, nodes in the OUT and IN are placed  between angles $\pi/2$- $5\pi /8$ and $3\pi/2$ - $13\pi /8$ respectively.  In addition, nodes in the OUT and IN are placed with an offset of 1.1 from the origin.  With the bow-tie representation we are able to visually compare the centrality of a node $i$ with node$j$.  Also, with this visualisation we are able to extract a network of nodes that mostly impact the others, nodes that impact just as much as they get impacted, and nodes that only get impacted by other nodes in the network.  The size and the colour of the node reflects impacted-impacting centrality of a node (nodes with larger impacted-impacting centrality are in red). The colour assigned to links is based on where the links point to in the network.  Links originating from IN to the SCC are in bright blue.  Links originating in the SCC to nodes in the SCC are in green.  Links that are originating in the SCC to the OUT are dull blue grey colour.}

\label{fig:bow-tie}
\end{centering}
\end{figure}
We then introduce a novel method for the visualisation of the bow-tie (Fig. \ref{fig:bow-tie}). 

This enables the representation, at the same time, of a network structure, the position of the nodes in the various component of the bow-tie, as well as their level of impacting centrality. In the diagram \ref{fig:bow-tie}, the circle represents the SCC, the top (bottom) section correspond to the IN (OUT). E.g. within the SCC, more central nodes are located towards the centre of the circle. The colour code and the size of the dots also reflect their centrality, such that the red and large dots are the most central (see caption of Fig. \ref{fig:bow-tie}). This visualisation allows to track how individual institutions become more or less central, or if they changed role across periods (see SI, Fig. 6 \& 7 b).

In period 1, most of the nodes of the bow-tie are in the SCC (85), with 4 and 6 in the IN and OUT respectively. Moreover, most nodes in the centre of the SCC are banks and investment banks, while insurance and real estate companies tend to be in the periphery of the SCC (see SI, Fig. 6 b).
 
This implies that in the normal phase most of the nodes impact the network, and are impacted by the network in a comparable manner. In period 2, the bow-tie grows overall, but the SCC (97 nodes) grows proportionally less than IN (19 nodes) and OUT (22 nodes). Of the 81 nodes that were disconnected in period 1, twenty seven migrated to the SCC (Fig. \ref{fig:bow-tie}). 
  
In period 3, the size of the bow-tie remains unchanged, but the SCC shrinks by about a 50\% (from 97 to 47 nodes), as a result of a migration to the IN (37 nodes) and mostly to the OUT (53 nodes) (see SI, Fig. 7 b).  In particular, the nodes with high impacting centrality are now all located in the IN and not, anymore, in the SCC. 

One should not forget that the original network is a strongly connected graph and the bow-tie is obtained with a filtering. Therefore, it is not the case that the nodes in the IN are not connected among each other. This means that in case a few nodes would have defaulted, the others would still have been heavily affected. However, the observed migration of nodes implies that, compared to the normal period, there has been an increasing polarisation between nodes (IN) that predominantly impact the network, and nodes (OUT) that predominantly are impacted by the network.

The above analysis of impacting and impacted centralities, and the bow-tie extraction allows us to move from an initial picture in which all nodes seemed to be equally important from the point of view of systemic risk, to a much more refined picture.  
 In terms of systemic impact, we can now focus on a small subset of the nodes, viz. nodes that have a high impacting centrality and are located in the centre of the SCC, or in the top part of the IN. This finding is corroborated by anecdotal evidence --\cite[see, e.g.,][]{fcic}--
%% TODO: fix. There is a way to include the e.g. in the parenthesis. Can't remember now.
 about the role of important actors of the credit crisis of 2008 (see SI). Conversely, the impacting centrality allows also to identify nodes that suffer the most from an impact originating from the others. Remarkably, there is no evidence of one or two nodes dominating the others in terms of systemic importance. In contrast, we see that in each period a set of about top 19 nodes have similar values of centrality.

\section{Discussion}
We have analysed the $\varepsilon$ drawup's in the CDS's time series for the top US and EU institutions throughout the last 10 years. By measuring the frequency of joint drawup's in pairs of CDS time series we have estimated the level of interdependence and trend reinforcement in the market. According to previous theoretical works on financial networks, the interplay of these two mechanisms is deeply linked to the emergence of systemic risk. We have found very significant levels of both, and an increase of both in acute phases of the crisis. The result suggests that high interdependence and trend reinforcement together with high level of individual riskiness are a good indicators of the level of systemic risk. Indeed, when CDS spreads where at their peak in 2008, implying high risk of individual default, movements in the spread of a few institutions were very likely to be followed by movements in another and {\it also in the same} institutions. This means that the default of a few players was likely to trigger a financial melt-down.

Furthermore, we have carried out what to our knowledge is the first study of the complex network of CDS interdependencies. In order to investigate the systemic importance of individual nodes, we have introduced two novel measures. The impacting centrality captures, in a recursive way, how much a node impacts the network. Symmetrically, the impacted centrality captures how much a node is impacted by the network. These two measures enable the extraction of a bow-tie structure from the initial network and to clarify the role of the nodes. We have found that in all phases of the market the top institutions by systemic importance are not just one or two, but about 19. 
% TODO: RK find a criteria/ check/ choose the right number
%% DONE: The number of guys that remain in the SCC across the three periods. 

They all have comparable level of systemic importance and they all are very tightly interconnected.

The specific findings of this analysis are relevant to the broad audience interested in the issue of systemic risk and systemically important financial institutions, including policy makers. 
Moreover, our approach is very general and applies to any set of time series associated to units that operate in interaction. In particular, it is of interest for those cases where the direct interaction between units is not observable and the dependence has to be inferred from the dynamics. In this respect, our paper contributes to a stream of work on the observability and the reconstruction of complex networks \citep{eldawlatly2010use},\citep{baca2009information},\citep{pfitzmann1987networks} .

\section{Methods}
\label{sec:methods}
\textbf{Data description}
The data consists of CDS spreads of single name entities denominated in US dollars and the Euro encompassing a total of 176 firms.  
\begin{figure}[htb!]
\begin{centering}
\includegraphics[width=0.97\textwidth]{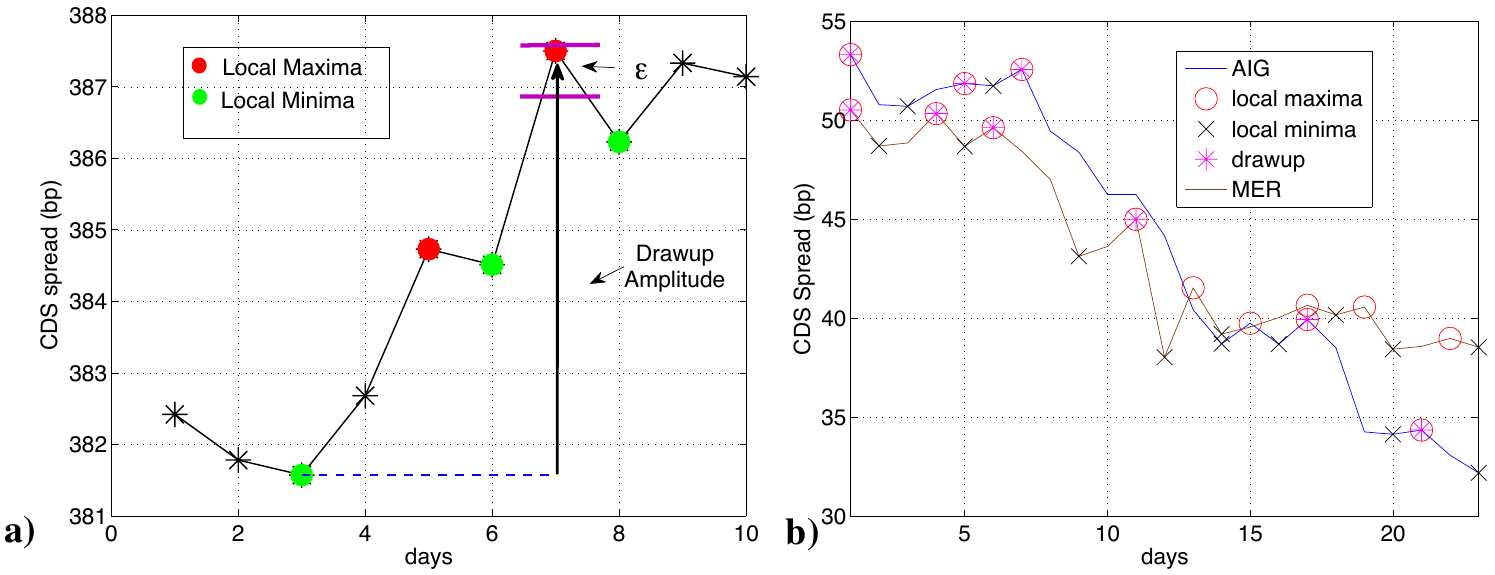} 
 \caption{ 
   \textbf{a)} Illustration of the $\varepsilon$-drawup methodology.  The \* represent local extrema that were not detected as drawups.  The red-dots represent local maxima that were picked up as candidates for a $\varepsilon$-drawup.  The green-dots represent the local minima that were picked up for a $\varepsilon$-drawup.  Compare the difference between the maxima and minima on days 5 and 6 respectively with $\varepsilon$.  Since, $\varepsilon$ > than the difference, we iterate to the next set of local maxima and minima on days 7 and 8, keeping day 3 as the day from when we count a $\varepsilon$-drawup
\textbf{b)} 
The plot highlights the $\varepsilon$-drawup methodology applied to the time series of American International Group (AIG) and Merrill Lynch (MER)
 }
 \label{fig:cds_traces_drawup_det}
\end{centering}
   
 \end{figure}

We have focused our attention on three different periods corresponding to different phases, see \ref{fig:traces}
of the market: (1) April 2003 - May 2006 (representative of a normal
phase); (2) May 2006 - March 2009 (volatile with an upwards trend);
(3) March 2009 - December 2011 (volatile with a downwards trend
market scenario).  The motivation to do a period-wise analysis is to extract clear snapshots of a network structure before, during and after the crisis of 2008.  This data window covers a period of 3624 trading days.   
%% TODO!!! 2010??? Or 2011?? Is this an old paragraph??? 
%% FIXED

\textbf{Data consistency}
The data was acquired via a subscription to Bloomberg.  The CDS data in general is provided to Bloomberg by various quote providers.  Bloomberg has negotiated contracts with various quote providers to display data on the Bloomberg terminal; however, not all quote providers relax the constraint of Bloomberg being allowed to distribute this data to the end user.  Thus, we see certain breaks in our time series, as on those days either the CDS was not traded (see SI), or is quoted by a quote provider that does not allow its data to be downloaded via Bloomberg.  To circumvent this issue, we carry forward the last traded price till a time that a new price has been reported. 

%\textbf{$\varepsilon$-drawup networks} 
% We dont mention anything about networks in this section thus dont see the point in calling this drawup networks.  Its only after the section on finite size that we are in a position to comment on what we interpret as an adjacency list.

\textbf{$\varepsilon$-drawups} We compute the $\varepsilon$-drawup's in each of the time series using the following algorithm, see figure \ref{fig:cds_traces_drawup_det}.  Suppose, we start our analysis of a $\varepsilon$-drawup from the first green point from the left in figure \ref{fig:cds_traces_drawup_det}b).  We follow the following steps to compute $\epsilon$-drawup's: (1) We compute the local variation in the time series for the last ten days, call it $\varepsilon$.  (2) compute local extrema.  (3) Goto first local minima, call it $\varepsilon$-$drawup_{candidate}$ (on day three, figure \ref{fig:cds_traces_drawup_det}a).  (4) Iterate to the set of local maxima and minima (occurring on days 5 and 6 respectively, see figure \ref{fig:cds_traces_drawup_det}b).  Compute the difference between the maxima and the minima, call it \textit{correction amplitude} (\textit{correction amplitude} refers to the decline in price followed after an increase in price).  (5) Update $\varepsilon$.  (6) If \textit{correction amplitude} $\geq$ $\epsilon$.  Then, we record the $\varepsilon$-drawup on the day it occurs (day 7 in figure \ref{fig:cds_traces_drawup_det}(b)).  And, we update  $\varepsilon$-$drawup_{candidate}$.  Otherwise, we iterate to the next minima and goto the succeeding maxima and repeat steps above.

%\begin{itemize}
%  \item compute local variation in the time series for the last ten days, call it $\varepsilon$
%  \item compute local extremas 
%  \item Goto first local minima, call $drawup_{candidate}$ (on day three, figure \ref{cds_traces_drawup_det}(b))
%  \item Iterate to the set of local maxima and minima (occurring on days 5 and 6 respectively, see figure \ref{fig:cds_traces_drawup_det}(b).  Compute the difference between the maxima and the minima, call it \textit{correction amplitude} (\textit{correction amplitude} refers to the decline in price followed after an increase in price)
%  \item Update $\varepsilon$
%  \item If \textit{correction amplitude} $\geq$ $\epsilon$ (.  Then,
%    \begin{itemize}
%      \item record drawup on the day it occurs (day 7 in figure \ref{cds_traces_drawup_det}(b))
%      \item update  $drawup_{candidate}$
%      \end{itemize}
%      \item Else: Iterate to next minima and goto succeeding maxima and repeat steps above
%    \end{itemize}
%

In this way, we are able to compute the $\varepsilon$-drawup's in the the time series data.  Once we have computed the $\varepsilon$-drawup's in the time series we map it into the vector $V_{i}$ of node $i$.  Where the length of the vector is the same as the length of the time series, $T$.  Also, $V_{i}(t) = 1$ if there was a drawup on day $t$, otherwise we put it to zero.  After proceeding with this methodology, we extract $N$ vectors of $\varepsilon$-drawup's, where $N$ is the number nodes in the analysis into a matrix of drawup's, $w_{i}$, where $i \in \{1, ..., N\}$.  

\textbf{Co-movements}
Once we have computed the matrix, $w_{i}$, we proceed to computing a matrix of co-movements.  We implement the following algorithm to compute co-movements:
(1) We select an arbitrary but fixed node $i$.  (2) We loop from $t = 1$ till $t = T$ and compare each $w_{i}(t)$ with all $w_{j}(t)$ where $j \in \{1,...,N\}$.  Compare $w_{i}(t)$ with all $w_{j}(t+\tau)$ where $j \in \{1,...,N\}$ and $\tau \in \{0,1,2,3\}$.  (3) If $w_{i}(t) = 1$ and $w_{j}(t) = 1$, then count = count $+ 1$.  If  If $w_{i}(t) = 1$ and $w_{j}(t+\tau) = 1$, then $count^{\tau} = count^{\tau} +1$, and then end the iteration.  (4) Update matrix of joint drawup's, i.e. $D_{ij} = \frac{count}{T}$, $D_{ij}^{\tau} = \frac{count^{\tau}}{T}$.  Repeat steps above till all the $i,j$ pairs have been accounted for.  We then filter each of the $\varepsilon$-drawup's based on a permutation test.  In order to estimate the probability that a node $j$ experiences a $\varepsilon$-drawup in the next three days given that node $i$ experiences a $\varepsilon$-drawup today, we sum the filtered matrices, i.e. $\sum_{\tau=0}^{\tau=3}D_{ij}^{\tau}$.  Notice that the quantity represented by $W_{ij}$ is not a measure of causality, as each $W_{ij}$ represents, in a probabilistic framework, the probability of co-movements and not that node $i$ causes a $\varepsilon$-drawup in node $j$.

%\begin{itemize}
%\label{item:algo2}
%\item Pick a node $i$ and fix it
%\begin{itemize}
%\item begin loop from $t = 1$ till $t = T$
%\item Compare each $w_{i}(t)$ with all $w_{j}(t)$ where $j \in \{1,...,N\}$.  Compare $w_{i}(t)$ with all $w_{j}(t+\tau)$ where $j \in \{1,...,N\}$ and $\tau \in \{0,1,2,3\}$
%\item If $w_{i}(t) = 1$ and $w_{j}(t) = 1$, then count = count $+ 1$.  If  If $w_{i}(t) = 1$ and $w_{j}(t+\tau) = 1$, then $count^{\tau} = count^{\tau} +1$.
%\item end loop
%\end{itemize}
%\item Update matrix of joint drawups, i.e. $D_{ij} = \frac{count}{T}$, $D_{ij}^{\tau} = \frac{count^{\tau}}{T}$.  
%\item repeat steps above till all pairs of $i,j$ are accounted for.
%\end{itemize}
%$j \in \{1,..., i-1, i+1,...,N\}$
The matrix of co-movements now carry probabilities of joint drawup's in the time series for all pairs.  The matrix of co-movements, constructed as above, is square; however, $D_{ij}, D_{ij}^{\tau} $ are not symmetric. 

\textbf{Randomness}.
To compute each element of the matrix $W_{ij}$ we subtract (element wise) from matrices $D_{ij}$ and $D_{ij}^{\tau}$ $D_{ii}*D_{jj}$ and $D_{ii}^{\tau}*D_{jj}^{\tau}$ respectively.  Alternatively, by the following procedure as above, we essentially conduct a test of independence, i.e. $W_{ij} = P_{ij} - P_{i}P_{j}$.  In addition, by construction $W_{ij}$ is adjusted for random occurrences.  

\textbf{Finite size}.
In order to clean the matrix of drawup's, $W_{ij}$.  
We permute the matrix indices of where the drawup's occur in $w_{i}$ .  We permute each pair of vectors $w_{i}$ and $w_{j}$ hundred times.  After each permutation we re-compute $W_{ij}^{control}(i,j)$ hundred times, following the procedure (see Methods), and extract a single $\tilde{W_{ij}}^{control}$ from hundred realisations at the 95\% confidence interval.  We then filter each element in $W_{ij}$ using our results from $\tilde{W_{ij}}^{control}$.  We then proceed to interpreting this filtered weighted matrix as an adjacency list of a directed network.

% \textbf{Centrality}.
% Once we have cleaned the empirical drawups matrix $W_{ij}$ we proceed to interpreting this matrix as an adjacency list.  Where each element $W_{ij}(i,j)$ is a weight assigned to $i$ influencing $j$.  

\textbf{Impacted and Impacted centrality}
In line with the notion of feedback centrality (e.g., PageRank see SI) we define the following measures. 
\begin{equation}
\label{eqn:b}
b_{i} = \sum_{j}\tilde{W}_{ij}v_{j}(0)+\beta\sum_{j}\tilde{W}_{ij}b_{j}
\end{equation}
\begin{equation}
\label{eqn:c}
c_{i} = \sum_{j}\tilde{W}^{'}_{ij}v_{j}(0)+\beta\sum_{j}\tilde{W}^{'}_{ij}c_{j}
\end{equation}
%b_{i} = \frac{\sum_{j}W_{ij}^{'}v_{j}(0)}{\sum_{l}W_{lj}{'}^{'}}+\frac{\sum_{j}W_{ij}^{'}b_{j}}{\sum_{l}W_{lj}^{'}}
%c_{i} = \frac{\sum_{j}W_{ij}v_{j}(0)}{\sum_{l}A_{jl}^{'}}+\frac{\sum_{j}W_{ij}^{'}c_{j}}{\sum_{l}A_{jl}}
where $\tilde{W}_{ij} = \frac{W_{ij}}{\sum_{l}W_{lj}}$, $v_{j}(0)=1$, and $\beta = 0.85$ is a dampening factor.  The matrix $\tilde{W}_{ij}$ is row stochastic and thus equations \ref{eqn:c} and \ref{eqn:b} can be analytically solved to yield the solution, $c= (\mathbb{I}- \mathbf{\tilde{W}}')^{-1}\mathbf{\tilde{W}}' v$ and $b= (\mathbb{I}- \mathbf{\tilde{W}})^{-1}\mathbf{\tilde{W}} b$.  Equation \ref{eqn:b} can be interpreted as the importance of a firm $i$ based on how much it gets impacted by the other firms in the system.  Equation \ref{eqn:c} can be interpreted as the centrality of a firm based on how it affects other members in the network.  Where $W_{ij}^{'}$ is the transpose of the adjacency matrix that we derive from the $\varepsilon$ drawup procedure and after having cleaned for randomness and finite size.  

The impacting centrality is analogous to: cumulatively how much of the distress in the network is due to node $i$.  Or, the loss experienced by the other nodes by the failure of node $i$.  Suppose there are $N$ shocks that are experienced by node $i$ that are propagated to the other nodes in the network based on $W_{ij}$.  Thus, by counting the number of times a shock from $i$ passes through the other nodes in the network represents the impact of node $i$ on the network.  Similarly, by counting the number of shocks in the entire system that propagate through node $i$ represent the impacted centrality of node $i$.

%%% Alternate to This is analogous to: ...1 PARAGRAPH TODO
The impacting centrality can be viewed as the cumulative sum of the shocks induced on all the nodes in the network cumulatively over time due to node $i$.  
 %TODO explin start with large nio. of shocks in node i.  let them travel according to Wij and count how many times a shock enters one of the other nodes.  establish parallel with websurfer (pagerank, random walkers)
OR, the expected no. of shocks that are propagated into the system given that I have a shock impacting centrality.

impacted centrality is the cumulative fraction of the shocks induced in me by a set of shocks on all the other nodes in the network over time.  %TODO explin start with large nio. of shocks in node i.  let them travel according to Wij and count how many times a shock enters one of the other nodes.  establish parallel with websurfer (pagerank, random walkers).  Assuing that in every node there are certain num. of shocks in all the other nodes (intitialy).  What is the fraction of these shocks (cumulativly) that enter my node.  no. of shocks that i get from other nodes from every possible walk (dicounted by beta).
OR, the expected no. of shocks that are propagated through me by all the other nodes in the system.

% To further derive an intuitive meaning from these centrality measures, we compute the ratio $\frac{b_{i}}{c_{i}}$.  In addition, if $\frac{b_{i}}{c_{i}} > 1.5$,  then, we retain all outgoing links from a firm to the other in the network, and truncate all the incoming links.  If $\frac{b_{i}}{c_{i}} > 0.75$, then we retain all the incoming links and truncate all the outgoing links.  Following this procedure, we, by construction, create a bow-tie structure.  In this fashion we are able to visually enhance the role of each of the market participants.  

\textbf{Bow-tie structure}
A bow-tie network is essentially a directed network consisting of four main bodies, namely a Strongly Connected Component (SCC); OUT: set of all nodes that can be reached, directly or indirectly, from the SCC;   IN: the set of all nodes that reach the SCC directly or indirectly.  The fourth and last component of the bow-tie structure, Tubes and Tendrils (TT) represent the set of all nodes that are not a part of the SCC; however, a node in the TT can either be reached from the IN and/or OUT.  If $r_{i} > 3/2$ then we remove all the incoming links.  This means that all nodes that exhibit $r_{i}>3/2$ only have outgoing links.  Similarly, for nodes that exhibit $r_{i}< 2/3$ have all the outgoing links removed.  And, the remaining nodes, i.e. $2/3 < r{i} < 3/2$ are in the SCC and retain both the incoming and outgoing links.  Clearly, this does not imply that nodes in the IN, or OUT do not have links among themselves; but, we simply disregard them by construction.   This merely highlights the nodes ability to impact its neighbours more than it gets affected and vice versa.  Thus, all nodes in the IN point to nodes in the SCS, and nodes in OUT get pointed to by the nodes in the SCC.

\subsection*{Acknowledgments}
\label{sec:acknowledgments}
The authors acknowledge the financial support from the Swiss National
Science Foundation Grant CR12I1-127000) and the European
Commission FET Open Project ``FOC'' 255987.

%\section*{References}
\bibliographystyle{apalike}
\bibliography{references_REF_CDSNETS11}

\end{document}